\documentclass[onecolumn]{article}

\usepackage[utf8]{inputenc}
\usepackage[T1]{fontenc}
\usepackage{amsmath,amssymb,amsfonts}
\usepackage{graphicx}
\usepackage{booktabs}
\usepackage[numbers]{natbib}
\usepackage[margin=1in]{geometry}
\usepackage{caption}
\usepackage{subcaption}
\usepackage{hyperref}
\usepackage{siunitx}
\usepackage{float}
\usepackage{xcolor}
\usepackage{multirow}

\title{\textbf{Sparse autoencoders reveal organized biological knowledge but minimal regulatory logic in single-cell foundation models: a comparative atlas of Geneformer and scGPT}}

\author{Ihor Kendiukhov\textsuperscript{1}\\[6pt]
\textsuperscript{1}Department of Computer Science, University of T\"ubingen, T\"ubingen, Germany\\
\texttt{kendiukhov@gmail.com}}

\date{}

\begin{document}

\maketitle

\begin{abstract}

\textbf{Background:}
Single-cell foundation models such as Geneformer and scGPT encode rich biological information, but whether this includes causal regulatory logic---as opposed to statistical co-expression---remains unclear. Sparse autoencoders (SAEs) can resolve superposition in neural networks by decomposing dense activations into interpretable features, yet have not been systematically applied to biological foundation models.

\textbf{Results:}
We trained TopK SAEs on residual stream activations from all layers of Geneformer V2-316M (18 layers, $d{=}1{,}152$) and scGPT whole-human (12 layers, $d{=}512$), producing atlases of 82,525 and 24,527 features, respectively. Both atlases confirm massive superposition: 99.8\% of features are invisible to SVD. Systematic characterization reveals rich biological organization: 29--59\% of features annotate to Gene Ontology, KEGG, Reactome, STRING, or TRRUST, with U-shaped layer profiles reflecting hierarchical abstraction. Features organize into co-activation modules (141 in Geneformer, 76 in scGPT), exhibit causal specificity (median 2.36$\times$), and form cross-layer information highways (63--99.8\%). However, when tested against genome-scale CRISPRi perturbation data, only 3/48 transcription factors (6.2\%) show regulatory-target-specific feature responses. A multi-tissue control yields marginal improvement (10.4\%, 5/48 TFs), establishing model representations as the bottleneck.

\textbf{Conclusions:}
These models have internalized organized biological knowledge---pathway membership, protein interactions, functional modules, hierarchical abstraction---but minimal causal regulatory logic. We release both feature atlases as interactive web platforms enabling exploration of over 107,000 features across 30 layers of two leading single-cell foundation models.
\end{abstract}


\section{Background}

Single-cell foundation models (scFMs) such as Geneformer~\cite{theodoris2023transfer} and scGPT~\cite{cui2024scgpt} have demonstrated remarkable capabilities across cell type annotation, perturbation response prediction, and gene network inference. Trained on millions of transcriptomic profiles, these models learn contextual gene representations that capture biological structure without explicit supervision on regulatory relationships. A central question for the field is whether these learned representations encode \emph{causal regulatory logic}---the directed relationships between transcription factors (TFs) and their target genes---or merely reflect \emph{statistical co-expression patterns} that correlate with but do not constitute regulation.

This question has been partially addressed at the level of attention weights. A companion study~\cite{kendiukhov2025systematic} systematically evaluated attention-derived edge scores from Geneformer and scGPT across 37 analyses and 153 statistical tests, finding that attention captures co-expression rather than unique regulatory signal: trivial gene-level baselines outperform attention edges, pairwise scores add zero incremental predictive value, and causal ablation of putatively regulatory heads produces no behavioural effect. However, attention weights represent only one view of a model's internal computation. The residual stream---the running sum of all layer outputs that carries information through the network---may encode richer structure than attention patterns alone reveal.

The superposition hypothesis~\cite{elhage2022superposition} provides a theoretical framework for understanding this structure. When a model must represent more concepts than it has dimensions, it encodes features as nearly-orthogonal directions in activation space, with any given input activating only a sparse subset. For Geneformer V2-316M with 1,152 hidden dimensions, superposition would allow the model to encode thousands of biological concepts---far more than the dimensionality suggests---in a manner invisible to standard linear decomposition methods like SVD or PCA.

Sparse autoencoders (SAEs)~\cite{olshausen1997sparse,sharkey2022taking,cunningham2023sparse,bricken2023monosemanticity} have emerged as a powerful tool for resolving superposition in neural networks. By learning an overcomplete dictionary with a sparsity constraint, SAEs decompose dense activations into interpretable features that correspond to meaningful concepts. This approach has yielded striking results in large language models, revealing interpretable features for safety-relevant concepts~\cite{templeton2024scaling} and scaling to billions of features~\cite{gao2024scaling}. However, SAEs have not been systematically applied to biological foundation models, where the ``concepts'' encoded are genes, pathways, and regulatory programs rather than linguistic entities.

Here, we present the first comprehensive SAE-based interpretability analysis of single-cell foundation models. We train TopK SAEs~\cite{makhzani2013ksparse} on per-position residual stream activations from all 18 layers of Geneformer V2-316M and all 12 layers of scGPT whole-human~\cite{cui2024scgpt}, producing atlases of 82,525 and 24,527 features, respectively. These two models differ substantially in architecture (rank-value vs.\ continuous-value encoding, 18 vs.\ 12 layers, $d{=}1{,}152$ vs.\ $d{=}512$) and training data (30M vs.\ 33M cells), yet we apply an identical analytical pipeline to both, enabling direct cross-model comparison.

We systematically characterize each atlas through complementary analyses organized in three phases: (1)~feature extraction, training, annotation, SVD comparison, and cross-layer tracking; (2)~co-activation network analysis, causal feature patching, cell type enrichment mapping, novel feature characterization, and cross-layer computational graph construction; and (3)~a multi-tissue control experiment testing whether the interpretability bottleneck lies in the SAE training data or in the model itself. To make these results accessible to the community, we release both atlases as interactive web platforms---the Geneformer Feature Atlas (\url{https://biodyn-ai.github.io/geneformer-atlas/}) and scGPT Feature Atlas (\url{https://biodyn-ai.github.io/scgpt-atlas/})---enabling exploration of over 107,000 features across 30 layers of two leading single-cell foundation models.

\section{Results}

\subsection{SAE feature atlas reveals massive superposition in Geneformer}
\label{sec:atlas}

We extracted per-position residual stream activations from all 18 layers of Geneformer V2-316M, processing 2,000 K562 control cells from the Replogle genome-scale CRISPRi dataset~\cite{replogle2022mapping} and obtaining 4,056,351 token positions per layer (mean 2,028 genes/cell; 336.4~GB total). For each layer, we trained a TopK sparse autoencoder with 4$\times$ overcomplete dictionary (4,608 features from 1,152 input dimensions) and $k{=}32$ sparsity, subsampling 1~million positions per layer for training efficiency (Methods).

\begin{figure}[t]
\centering
\includegraphics[width=\textwidth]{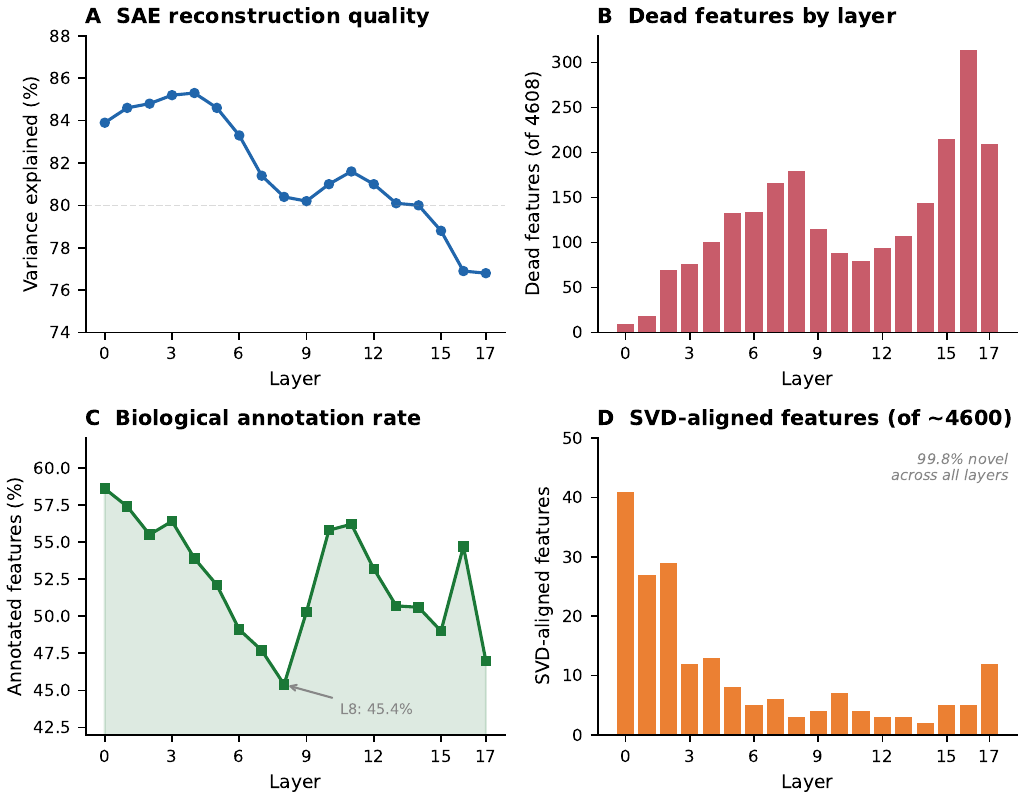}
\caption{\textbf{Geneformer V2-316M SAE feature atlas.} TopK SAEs on all 18 layers yield 82,525 features. Reconstruction quality declines with depth while dead features increase.}
\label{fig:overview}
\end{figure}

Table~\ref{tab:full18} presents complete training and annotation statistics for all 18 layers (Fig.~\ref{fig:overview}). Several patterns emerge. Variance explained peaks at layers 3--4 (85.2--85.3\%) and generally declines thereafter to 76.8\% at layer~17 (with a brief recovery at layers 10--11), indicating that later-layer representations become progressively more distributed and harder to compress with a fixed-size dictionary. Dead features (those never activating on 100K held-out positions) increase from 0 at layer~0 to 70 at layer~16, with a trough at layers 9--11 (6--13 dead) suggesting a ``mid-layer revival'' of structured representations. Feature orthogonality is excellent throughout: mean absolute inter-feature cosine similarity ranges from 0.033--0.040, indicating well-separated dictionary directions.

The total atlas comprises \textbf{82,525 alive features} across all 18 layers, with \textbf{43,241 features} (52.4\%) receiving at least one significant ontology annotation against five biological databases (Gene Ontology Biological Process~\cite{ashburner2000go}, KEGG~\cite{kanehisa2000kegg}, Reactome~\cite{jassal2020reactome}, STRING~\cite{szklarczyk2023string}, and TRRUST~\cite{han2018trrust}).

\begin{table}[t]
\centering
\caption{\textbf{Geneformer SAE training and annotation statistics.} Each SAE uses a 4$\times$ overcomplete dictionary (4,608 features) with TopK $k{=}32$ sparsity. VarExpl = variance explained. Dead = $4{,}608 - \text{Alive}$. MeanCos = mean absolute pairwise cosine between decoder vectors. Ann\% = fraction of alive features with $\geq$1 significant enrichment (FDR $< 0.05$). Enrichments = unique (feature, term) pairs. Per-database breakdown in Additional file~1: Table~S1.}
\label{tab:full18}
\smallskip
\begin{tabular}{rccrcrccc}
\toprule
Layer & VarExpl & Alive & Dead & SVD & Novel & MeanCos & Ann\% & Enrichments \\
\midrule
0  & 83.9\% & 4,608 & 0   & 41 & 4,567 & 0.033 & 58.6\% & 23,959 \\
1  & 84.6\% & 4,606 & 2   & 27 & 4,579 & 0.033 & 57.4\% & 23,131 \\
2  & 84.8\% & 4,601 & 7   & 29 & 4,572 & 0.034 & 55.5\% & 23,383 \\
3  & 85.2\% & 4,595 & 13  & 12 & 4,583 & 0.035 & 56.4\% & 21,922 \\
4  & 85.3\% & 4,583 & 25  & 13 & 4,570 & 0.037 & 53.9\% & 19,910 \\
5  & 84.6\% & 4,576 & 32  & 8  & 4,568 & 0.036 & 52.1\% & 17,865 \\
6  & 83.3\% & 4,580 & 28  & 5  & 4,575 & 0.035 & 49.1\% & 16,716 \\
7  & 81.4\% & 4,591 & 17  & 6  & 4,585 & 0.036 & 47.7\% & 15,470 \\
8  & 80.4\% & 4,586 & 22  & 3  & 4,583 & 0.038 & 45.4\% & 15,679 \\
9  & 80.2\% & 4,595 & 13  & 4  & 4,591 & 0.036 & 50.3\% & 16,925 \\
10 & 81.0\% & 4,602 & 6   & 7  & 4,595 & 0.035 & 55.8\% & 19,587 \\
11 & 81.6\% & 4,598 & 10  & 4  & 4,594 & 0.037 & 56.2\% & 19,943 \\
12 & 81.0\% & 4,592 & 16  & 3  & 4,589 & 0.037 & 53.2\% & 19,105 \\
13 & 80.1\% & 4,583 & 25  & 3  & 4,580 & 0.038 & 50.7\% & 17,338 \\
14 & 80.0\% & 4,568 & 40  & 2  & 4,566 & 0.040 & 50.6\% & 17,719 \\
15 & 78.8\% & 4,543 & 65  & 5  & 4,538 & 0.038 & 49.0\% & 15,571 \\
16 & 76.9\% & 4,538 & 70  & 5  & 4,533 & 0.037 & 54.7\% & 16,080 \\
17 & 76.8\% & 4,580 & 28  & 12 & 4,568 & 0.039 & 47.0\% & 15,764 \\
\midrule
\textbf{Total} & & \textbf{82,525} & \textbf{419} & \textbf{189} & \textbf{82,336} & & & \textbf{336,067} \\
\bottomrule
\end{tabular}
\end{table}

\subsection{Quantifying superposition: SAE features versus SVD}
\label{sec:svd}

To quantify the degree of superposition, we systematically compared SAE features against the top-50 SVD axes at each layer. A feature was classified as ``SVD-aligned'' if its decoder weight vector had cosine similarity $> 0.7$ with any SVD axis.

Across all 18 layers, only \textbf{189 of 82,525} features (0.2\%) aligned with any SVD axis. SVD-aligned features decrease sharply with depth: 41 at layer~0, declining to just 2--5 at mid-to-late layers (Table~\ref{tab:full18}). The remaining 99.8\% (82,336 features) represent structure in the residual stream that is invisible to standard linear decomposition.

Critically, the novel features carry the biological signal:
\begin{itemize}
\item \textbf{Annotation rate:} 52.5\% of novel features have ontology annotations versus only 14.3\% of SVD-aligned features.
\item \textbf{Absolute counts:} 43,214 novel features are annotated versus 27 SVD-aligned.
\item \textbf{Exclusivity:} 98.7\% of all ontology enrichment terms are found \emph{exclusively} in novel features.
\item \textbf{Variance:} SAEs explain 77--85\% of activation variance versus 31--38\% for the top-50 SVD axes (2.4$\times$ ratio).
\end{itemize}

\begin{figure}[t]
\centering
\includegraphics[width=\textwidth]{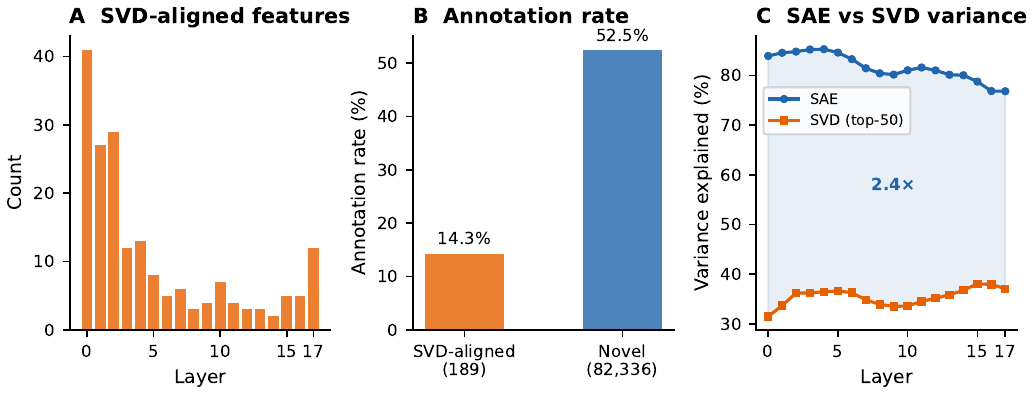}
\caption{\textbf{99.8\% of SAE features are invisible to SVD.} Novel features carry 98.7\% of annotations and explain 2.4$\times$ more variance.}
\label{fig:svd}
\end{figure}

These results confirm that Geneformer uses massive superposition to encode its biological knowledge (Fig.~\ref{fig:svd}): the model represents at least 82,525 biological concepts in 1,152 dimensions---a compression ratio exceeding 70$\times$.

\subsection{scGPT feature atlas: cross-model replication}
\label{sec:scgpt_atlas}

To test whether the superposition and biological organization observed in Geneformer generalize across architectures, we applied an identical SAE pipeline to scGPT whole-human~\cite{cui2024scgpt}---a model trained on 33 million cells with fundamentally different design choices: continuous-value gene encoding (vs.\ Geneformer's rank-value tokens), 12 transformer layers (vs.\ 18), $d_\text{model}{=}512$ (vs.\ 1,152), and masked gene prediction training objective (vs.\ next-token prediction).

We extracted per-position residual stream activations from all 12 layers, processing 3,000 diverse Tabula Sapiens cells (1,000 immune across 43 cell types, 1,000 kidney, 1,000 lung) and obtaining 3,561,832 token positions per layer. For each layer, we trained a TopK SAE with 4$\times$ overcomplete dictionary (2,048 features from 512 input dimensions) and $k{=}32$ sparsity.

Table~\ref{tab:scgpt_full12} presents complete training statistics for all 12 scGPT layers. Reconstruction quality is notably higher than Geneformer: variance explained ranges from 85.7\% (L7) to 93.5\% (L4), with a mean of 90.2\% versus Geneformer's mean of 81.7\%. Dead features are near-zero: only 49 across all 12 layers (0.2\%), compared to 419 in Geneformer (0.5\%). The total atlas comprises \textbf{24,527 alive features}, with \textbf{7,595 features} (31.0\%) annotated against the same five ontology databases.

\begin{table}[t]
\centering
\caption{\textbf{scGPT SAE training and annotation statistics.} 4$\times$ dictionary (2,048 features), TopK $k{=}32$, 3.56M positions/layer. Notation as in Table~\ref{tab:full18}.}
\label{tab:scgpt_full12}
\smallskip
\begin{tabular}{rccrccc}
\toprule
Layer & VarExpl & Alive & Dead & MeanCos & Ann\% & Modules \\
\midrule
0  & 92.0\% & 2,027 & 21 & 0.038 & 32.7\% & 6 \\
1  & 93.0\% & 2,035 & 13 & 0.041 & 30.8\% & 6 \\
2  & 93.2\% & 2,038 & 10 & 0.040 & 28.7\% & 5 \\
3  & 93.2\% & 2,045 & 3  & 0.041 & 31.6\% & 7 \\
4  & 93.5\% & 2,048 & 0  & 0.041 & 30.4\% & 6 \\
5  & 92.1\% & 2,048 & 0  & 0.042 & 29.6\% & 7 \\
6  & 90.3\% & 2,048 & 0  & 0.046 & 28.9\% & 7 \\
7  & 85.7\% & 2,048 & 0  & 0.049 & 32.4\% & 6 \\
8  & 86.3\% & 2,048 & 0  & 0.046 & 28.9\% & 7 \\
9  & 86.9\% & 2,047 & 1  & 0.045 & 33.9\% & 5 \\
10 & 87.4\% & 2,047 & 1  & 0.044 & 31.7\% & 7 \\
11 & 88.6\% & 2,048 & 0  & 0.043 & 32.0\% & 7 \\
\midrule
\textbf{Total} & & \textbf{24,527} & \textbf{49} & & & \textbf{76} \\
\bottomrule
\end{tabular}
\end{table}

\begin{figure}[t]
\centering
\includegraphics[width=\textwidth]{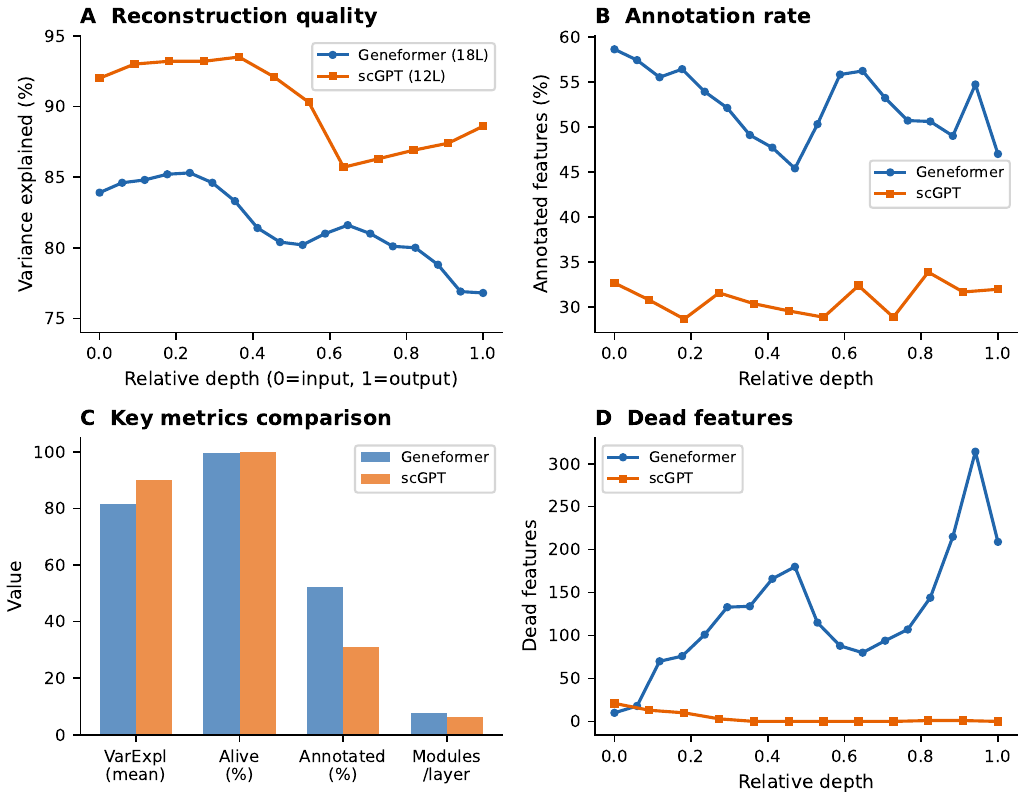}
\caption{\textbf{Cross-model comparison on normalized depth.} (A)~scGPT has higher variance explained at all depths. (B)~Geneformer has higher annotation rates. (C)~Summary. (D)~Dead feature profiles differ.}
\label{fig:cross_model}
\end{figure}

Table~\ref{tab:model_comparison} provides a direct head-to-head comparison (Fig.~\ref{fig:cross_model}). Despite their architectural differences, both models exhibit the same qualitative phenomena: massive superposition, high reconstruction quality, rich biological annotation, organized modular structure, and cross-layer information highways. Key quantitative differences include scGPT's higher variance explained (90.2\% vs.\ 81.7\%) and lower annotation rate (31.0\% vs.\ 52.4\%). The higher variance explained likely reflects the lower dimensionality making the 4$\times$ expansion a less extreme overcomplete basis, while the lower annotation rate may result from the continuous-value encoding distributing information more evenly across features.

\begin{table}[t]
\centering
\caption{\textbf{Cross-model comparison.} Both models analyzed with identical TopK SAE pipeline (4$\times$ expansion, $k{=}32$).}
\label{tab:model_comparison}
\smallskip
\begin{tabular}{lcc}
\toprule
Metric & Geneformer & scGPT \\
\midrule
Architecture & 18 layers, $d{=}1{,}152$ & 12 layers, $d{=}512$ \\
Training data & $\sim$30M cells & $\sim$33M cells \\
Input encoding & Rank-value tokens & Continuous expression \\
Training objective & Next-token prediction & Masked gene prediction \\
\midrule
Features per layer & 4,608 & 2,048 \\
Total alive features & 82,525 & 24,527 \\
Dead features & 419 (0.5\%) & 49 (0.2\%) \\
Mean VarExpl & 81.7\% & 90.2\% \\
Mean annotation rate & 52.4\% & 31.0\% \\
Total modules & 141 & 76 \\
Mean modules/layer & 7.8 & 6.3 \\
Module coverage & 96.0--99.5\% & 96.3\% mean \\
\bottomrule
\end{tabular}
\end{table}

\subsection{Biological annotation reveals a U-shaped layer profile}
\label{sec:ushape}

\begin{figure}[t]
\centering
\includegraphics[width=\textwidth]{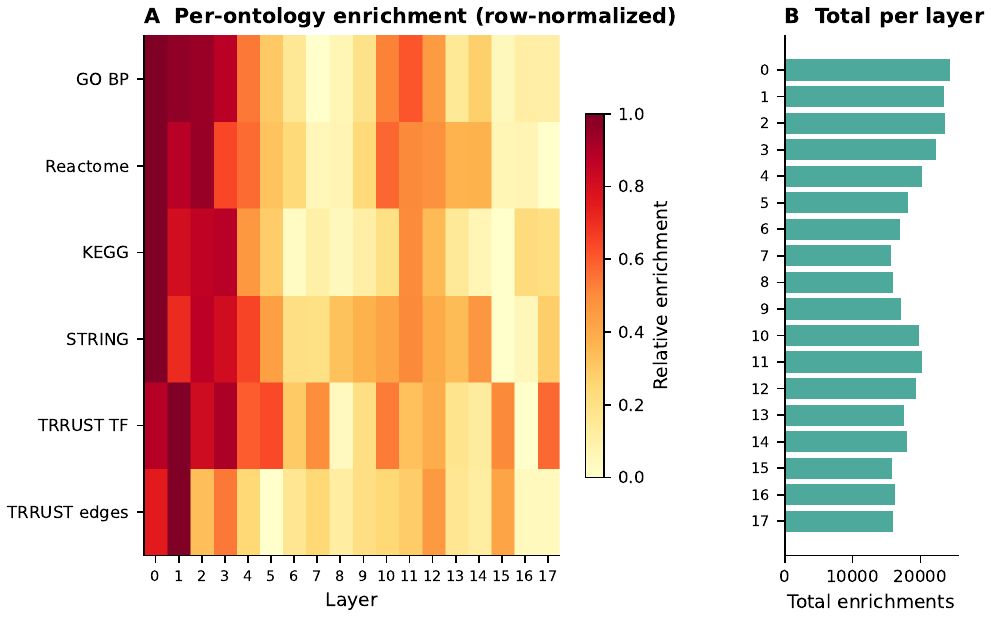}
\caption{\textbf{U-shaped annotation profile across 18 layers.} Early layers encode molecular programs; middle layers develop abstract representations; later layers re-specialize then optimize for prediction.}
\label{fig:ushape}
\end{figure}

Annotation of each feature's top-20 genes against five databases (Fisher's exact test, Benjamini--Hochberg FDR $< 0.05$) revealed a striking U-shaped profile across layers (Fig.~\ref{fig:ushape}). Annotation rates are highest at layers 0--1 (57--59\%), decline to a minimum at layer~8 (45.4\%), recover to 55--56\% at layers 10--11, and decrease again at layers 16--17 (47--55\%). Per-ontology counts are detailed in Additional file~1: Table~S1.

The scGPT atlas shows a weaker but qualitatively similar pattern across its 12 layers (Fig.~\ref{fig:cross_model}): annotation rates range from 28.7\% (L2) to 33.9\% (L9). Lower annotation rates cluster at L2 (28.7\%), L5 (29.6\%), L6 (28.9\%), and L8 (28.9\%), interspersed with higher rates at L7 (32.4\%) and L9--L11 (31.7--33.9\%).

For Geneformer, this pattern admits a functional interpretation organized into four zones:

\paragraph{Early layers (0--4): Molecular machinery.} Features map cleanly onto existing ontology terms, with the highest GO~BP (8,500--10,150 enrichments/layer), KEGG (2,050--2,650), and Reactome (9,200--11,000) counts. STRING protein--protein interaction associations peak here (248--302 per layer), as do TRRUST TF associations (133--164). Representative features encode textbook biological programs (Table~\ref{tab:examples}).

\paragraph{Middle layers (5--9): Abstract computation.} Annotation rates drop below 50\% at layers 6--8, consistent with intermediate computational representations harder to map to single ontology terms. GO~BP drops to 6,600--7,700, and STRING associations decrease to 181--216 per layer.

\paragraph{Mid-late layers (10--12): Re-specialization.} Annotation rates recover to 53--56\%, with GO~BP returning to 8,200--8,800 and KEGG to 1,750--2,100. Features shift from molecular processes to integrative cellular programs such as cell differentiation, intracellular signaling, and organelle organization.

\paragraph{Terminal layers (15--17): Prediction-focused.} A second annotation decline (47--55\%), with the most dead features (28--70) and most distributed representations. Features at these layers respond broadly to perturbations (Section~\ref{sec:perturbation}) but lack the specificity of mid-layer features.

\begin{table}[t]
\centering
\caption{\textbf{Representative SAE features.} Top genes = highest mean activation magnitude. Databases = ontology sources with significant enrichments.}
\label{tab:examples}
\smallskip
\small
\begin{tabular}{cp{3.8cm}p{3.5cm}l}
\toprule
Feature & Top Genes & Biological Identity & Databases \\
\midrule
\multicolumn{4}{l}{\textit{Layer 0 --- Molecular Programs}} \\
F3717 & CDK1, CDC20, DLGAP5, PBK & Cell cycle (G2/M) & GO, KEGG, React., STRING \\
F3607 & RRM1, E2F1, MCM4, MCM6, RAD51 & DNA replication / repair & GO, KEGG, React., STRING, TRRUST \\
F1116 & ARL6IP4, SQSTM1, JAK1 & B cell activation & GO, KEGG, Reactome \\
F4536 & DYNC1H1, TLN1, MYH9, SPTAN1 & Cytoskeleton / focal adhesion & GO, KEGG, Reactome \\
F2829 & TGFB1, PKN1, GADD45A, ZBTB7A & MAPK/TGF$\beta$ signaling & GO, KEGG, React., STRING, TRRUST \\
F1573 & MKI67, CENPF, TOP2A, CDK1, AURKB & Mitosis / chr.\ segregation & GO, KEGG, React., STRING \\
\midrule
\multicolumn{4}{l}{\textit{Layer 11 --- Integrative Programs}} \\
F2035 & \textit{(cell diff.\ genes)} & Cell diff.\ (neg.\ reg.) & GO, Reactome \\
F3692 & \textit{(ERAD pathway genes)} & ER-associated degradation & GO, KEGG, Reactome \\
F3933 & \textit{(signaling genes)} & Intracell.\ signaling (neg.\ reg.) & GO, Reactome \\
F2936 & \textit{(mitochondrial genes)} & Mitochondrion organization & GO, KEGG, Reactome \\
\bottomrule
\end{tabular}
\end{table}

\subsection{Cross-layer tracking: features are layer-specific}
\label{sec:crosslayer}

To understand how features relate across layers, we tracked decoder weight vector similarity from layer~0 features to all subsequent layers (Additional file~1: Table~S2). Features are overwhelmingly layer-specific. Only 2--3\% of features at any layer match a feature at the adjacent layer. From layer~0, the decay is steep: 114 matches at L1 (2.5\%), 67 at L4 (1.5\%), 10 at L8 (0.2\%), and effectively zero beyond L10. No layer~0 feature survives past layer~11.

A striking finding is that feature persistence anti-correlates with biological content. Of layer~0 features, 98.2\% are transient (present in $\leq$3 layers), with 59.6\% annotated and a mean of 5.4 enrichment terms. The rare moderate-persistence features (1.8\%, present in 4--10 layers) have only 3.7\% annotation rate and 0.1 mean enrichments. The biological content is carried by features rebuilt at every layer, not by any persistent scaffold.

\subsection{Features organize into 141 biologically coherent co-activation modules}
\label{sec:modules}

To map the relational structure among features, we constructed co-activation graphs at each of the 18 layers using pointwise mutual information (PMI~\cite{manning1999foundations}) computed over the 4~million token positions. Feature $i$ is ``active'' at position $j$ if it is among the top-$k$ activations for that position. PMI quantifies whether two features co-activate more often than expected: $\text{PMI}(i,j) = \log_2 P(i,j) / [P(i)P(j)]$. Leiden community detection~\cite{traag2019louvain} at resolution~1.0 identified co-activation modules.

Across all layers, we identified \textbf{141 distinct modules} (6--12 per layer) covering \textbf{96.0--99.5\%} of alive features (Additional file~1: Table~S3). This is not a trivial result: with $k{=}32$ sparsity out of 4,608 features, random co-activation would produce a connected but undifferentiated graph. Instead, Leiden clustering identifies well-separated communities with distinct biological identity.

\begin{figure}[t]
\centering
\includegraphics[width=\textwidth]{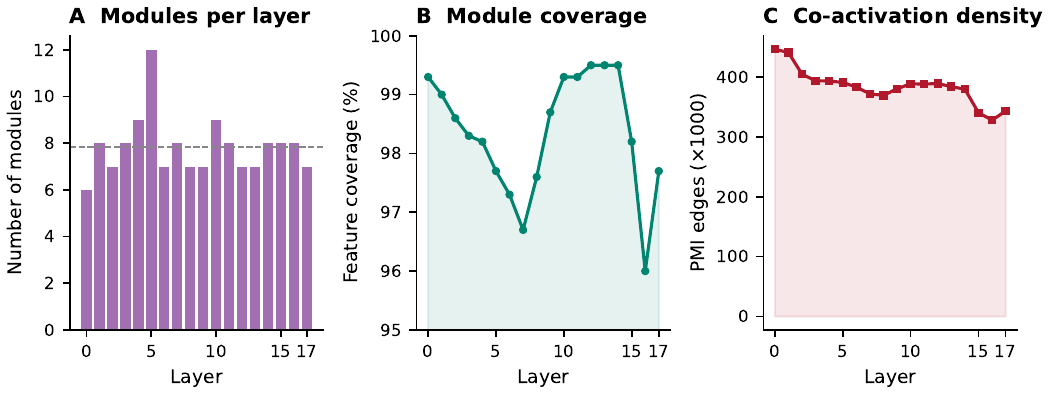}
\caption{\textbf{Co-activation modules evolve across layers.} Identity shifts from molecular machinery (L0) to integrative programs (L11), reflecting hierarchical abstraction.}
\label{fig:modules}
\end{figure}

Several structural patterns emerge (Fig.~\ref{fig:modules}). Module count peaks at layer~5 (12 modules), suggesting maximal functional compartmentalization at early-to-mid processing. PMI edge density declines with depth (446K at L0 to 328K at L16), paralleling the declining variance explained.

Modules have clear biological identity. At layer~0, six modules correspond to: (1)~Cell Cycle / DNA Replication, (2)~Immune Signaling, (3)~Metabolism, (4)~Translation, (5)~Protein Quality Control, and (6)~Cytoskeleton / Adhesion. At layer~11, eight modules shift toward integrative cellular programs: Cell Differentiation, Intracellular Signaling, Mitochondrial Organization, Vesicular Transport, Chromatin/Transcription, Protein Modification, Cell Motility, and Stress Response.

The scGPT atlas exhibits the same modular organization: 76 modules across 12 layers (5--7 per layer), with mean 96.3\% feature coverage. Despite the smaller dictionary size (2,048 vs.\ 4,608), module count per layer is comparable (6.3 vs.\ 7.8 mean), suggesting that the number of biological programs discoverable by SAEs scales sub-linearly with dictionary size.

\subsection{Causal patching demonstrates feature-level specificity}
\label{sec:causal}

Co-activation and annotation establish correlation between features and biology, but not causation. To test whether individual SAE features are causally necessary for the model's computations, we performed causal feature patching at layer~11. For each of 50 richly annotated features, we zeroed its SAE activation in the hidden state (via a forward hook), continued the forward pass through layers 12--17, and measured the resulting change in output logits. Specificity was quantified as the ratio of logit disruption at gene positions matching the feature's ontology annotation (``target genes'') to disruption at all other positions (``other genes'').

\begin{figure}[t]
\centering
\includegraphics[width=\textwidth]{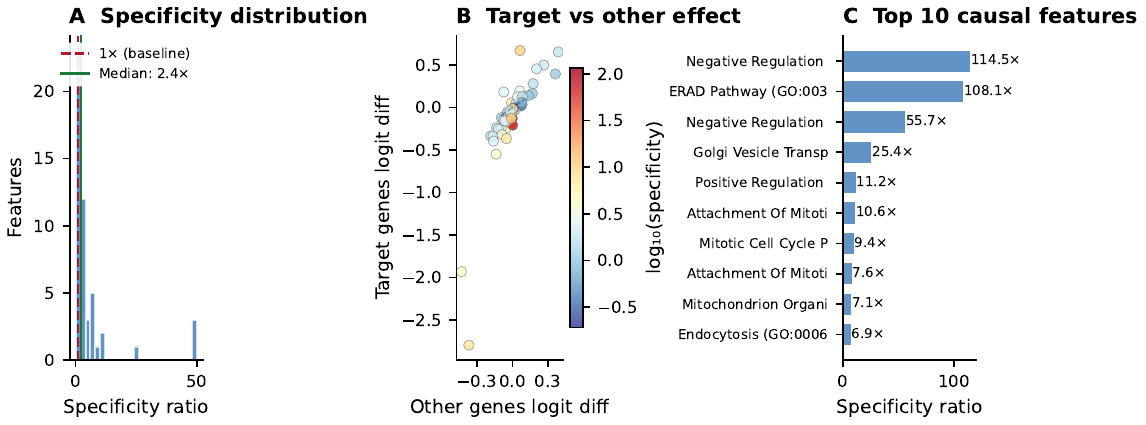}
\caption{\textbf{SAE features are causally necessary.} Ablating a single feature specifically disrupts its annotated gene set. Median 2.36$\times$; top feature 114.5$\times$.}
\label{fig:causal}
\end{figure}

Table~\ref{tab:causal_summary} summarizes the results (Fig.~\ref{fig:causal}). The median specificity ratio is \textbf{2.36$\times$}, with \textbf{60\%} of features exceeding 2$\times$ and \textbf{12\%} exceeding 10$\times$. The mean target logit disruption ($-0.116$) is 23$\times$ larger than the mean off-target disruption ($-0.005$), confirming that feature ablation preferentially affects the feature's annotated biology. The ten most specific features are detailed in Additional file~1: Table~S4; the top feature (F2035, negative regulation of cell differentiation) achieves 114.5$\times$ specificity.

\begin{table}[t]
\centering
\caption{\textbf{Causal patching summary.} 50 features at layer~11, 200 cells each. Specificity = $|\Delta\text{logit}_\text{target}| / |\Delta\text{logit}_\text{other}|$.}
\label{tab:causal_summary}
\smallskip
\begin{tabular}{lr}
\toprule
Metric & Value \\
\midrule
Features tested & 50 \\
Mean specificity ratio & 8.98 \\
Median specificity ratio & 2.36 \\
Specific ($>$2$\times$) & 30/50 (60\%) \\
Highly specific ($>$10$\times$) & 6/50 (12\%) \\
Mean target logit diff & $-$0.116 \\
Mean other logit diff & $-$0.005 \\
\bottomrule
\end{tabular}
\end{table}

This contrasts sharply with the companion study's finding~\cite{kendiukhov2025systematic} that ablating entire attention heads or MLP layers at the component level produces null behavioral effects. Feature-level interventions are sufficiently targeted to reveal genuine computational structure that coarser component-level interventions miss.

\paragraph{scGPT causal patching at layer~7.} We performed the same experiment on scGPT at layer~7 (approximately two-thirds through each stack). The median specificity was 0.98$\times$ (mean 1.02$\times$), with 0/50 features exceeding 2$\times$ specificity. This substantially weaker causal signal likely reflects the use of proxy expression values (uniform 1.0 for all genes) rather than actual expression levels, because the original continuous expression values were not preserved during activation extraction. The computational graph analysis (Section~\ref{sec:highways}) confirms that scGPT features are genuinely interconnected.

\subsection{Cross-layer information highways}
\label{sec:highways}

Given that features are almost entirely layer-specific (Section~\ref{sec:crosslayer}), how does biological information flow through the network? We computed cross-layer PMI between SAE feature activations at three layer pairs (L0$\to$L5, L5$\to$L11, L11$\to$L17), encoding the same 500,000 positions through both layer's SAEs (Additional file~1: Table~S5). An ``information highway'' is a source-layer feature with at least one strong (PMI $> 3$) dependency on a target-layer feature.

\begin{figure}[t]
\centering
\includegraphics[width=\textwidth]{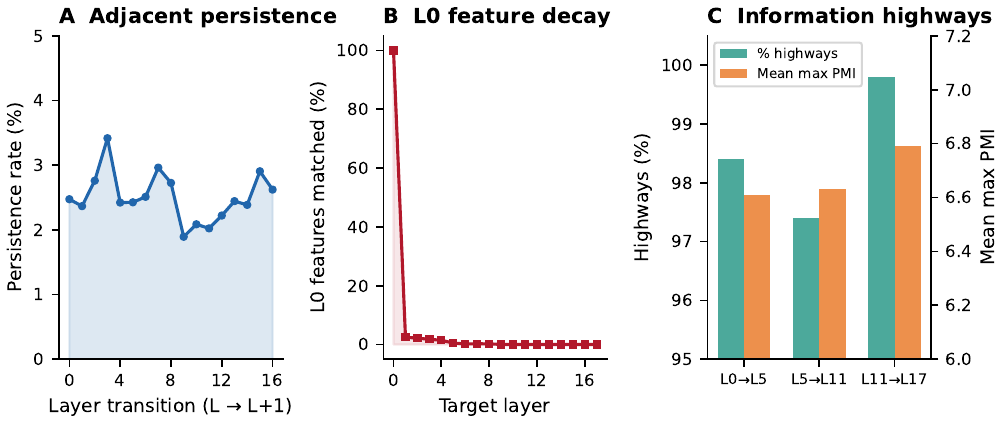}
\caption{\textbf{97--99.8\% of features form information highways.} Biological information flows through functional connections between distinct feature sets across layers.}
\label{fig:highways}
\end{figure}

Despite near-zero decoder-weight similarity across layers, functional connectivity is pervasive (Fig.~\ref{fig:highways}). Between \textbf{97.4\%} and \textbf{99.8\%} of features at each layer are information highways. Mean maximum PMI ranges from 6.61 to 6.79, with individual connections reaching PMI $> 10$. The L11$\to$L17 transition shows the densest connectivity (99.8\%), suggesting that information flow \emph{increases} toward the output.

Biologically meaningful cascades emerge among the strongest cross-layer connections (Additional file~1: Table~S7). The mTORC1~regulation $\to$ autophagy connection (L0$\to$L5, PMI~=~9.55) recapitulates a well-known signaling axis. Protein modification features at L11 connect to angiogenesis regulation at L17 (PMI~=~10.62).

\paragraph{scGPT cross-layer graph.} scGPT reveals a progressive concentration pattern (Additional file~1: Table~S6, Fig.~S1): upstream connectivity is consistently high (86.6--95.5\%), but downstream connectivity drops from 95.7\% (L0$\to$L4) to 62.9\% (L8$\to$L11). This suggests that scGPT progressively bottlenecks information toward later layers---a pattern not observed in Geneformer, where downstream connectivity remains near-complete (97.4--99.8\%).

\subsection{Perturbation response mapping reveals detection without regulatory specificity}
\label{sec:perturbation}

The preceding sections establish that SAE features are biologically annotated, modularly organized, causally specific, and functionally connected. We now ask the critical question: does this atlas encode \emph{regulatory logic}?

We mapped perturbation responses using CRISPRi knockdown cells from the Replogle dataset~\cite{replogle2022mapping}: 100 targets (48 TRRUST TFs and 52 other genes), 20 perturbed cells per target, compared against a control baseline of 100K positions. A feature ``responds'' if its activation changes significantly (Wilcoxon test, FDR~$< 0.05$, $|\text{effect}| > 0.5$). For each TRRUST TF, we tested whether responding features are enriched for the TF's known regulatory targets.

\begin{figure}[t]
\centering
\includegraphics[width=\textwidth]{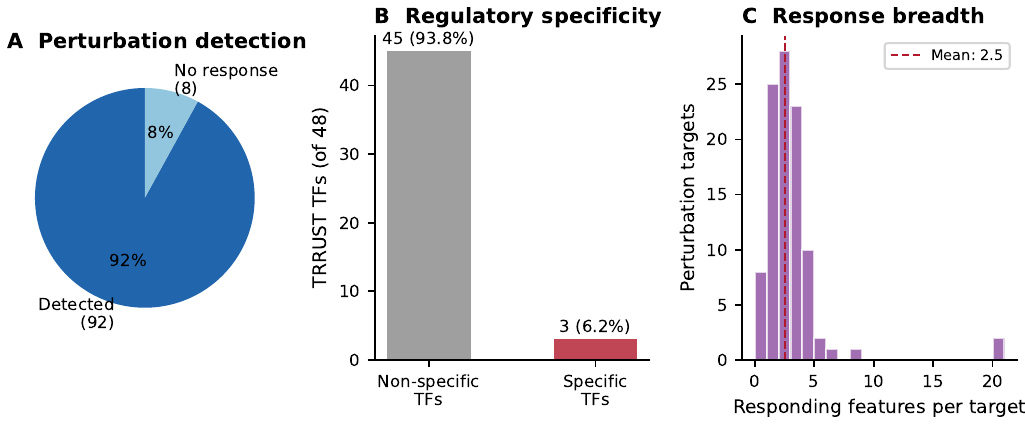}
\caption{\textbf{92\% perturbation detection but only 6.2\% TF specificity.} Features respond to cell-state shifts, not specific regulatory consequences.}
\label{fig:perturbation}
\end{figure}

Table~\ref{tab:perturbation} summarizes the results (Fig.~\ref{fig:perturbation}). The model detects perturbations: \textbf{92\%} of knockdowns cause at least one significant SAE feature change, with a mean of 2.54 responding features per target. However, the responses are overwhelmingly non-specific: only \textbf{3 of 48 TRRUST TFs (6.2\%)} produce feature responses that match their known regulatory targets.

\begin{table}[t]
\centering
\caption{\textbf{Perturbation response mapping (K562-only SAE, layer~11).} 100 CRISPRi targets tested. TF specificity = fraction of TFs with target-specific responding features.}
\label{tab:perturbation}
\smallskip
\begin{tabular}{lr}
\toprule
Metric & Value \\
\midrule
Perturbation targets tested & 100 \\
Targets causing feature changes & 92/100 (92\%) \\
TRRUST TFs tested & 48 \\
TFs with specific response & 3/48 (6.2\%) \\
Mean responding features per target & 2.54 \\
Mean specific features per target & 0.03 \\
\bottomrule
\end{tabular}
\end{table}

This 6.2\% specificity rate is the central negative finding (Fig.~\ref{fig:perturbation}). The model knows \emph{that} a perturbation has occurred---it detects the shift in cell state---but does not encode \emph{which specific regulatory targets} should be affected. This confirms, at the granular SAE feature level, what the companion attention-based analysis found~\cite{kendiukhov2025systematic}: Geneformer's internal representations encode co-expression structure and pathway membership, but not the directed TF$\to$target regulatory wiring.

\subsection{Multi-tissue SAE confirms the model as the bottleneck}
\label{sec:multitissue}

The low perturbation specificity (6.2\%) has two possible explanations: (a)~a limitation of the SAE training data---K562 cells lack TF diversity; or (b)~a limitation of Geneformer itself. To distinguish these, we trained multi-tissue SAEs on pooled activations from K562 and diverse Tabula Sapiens~\cite{tabula2022tabula} cells (3,000 cells: 1,000 immune across 43 cell types, 1,000 kidney, 1,000 lung; 500K K562 and 500K Tabula Sapiens positions per layer).

\begin{figure}[t]
\centering
\includegraphics[width=\textwidth]{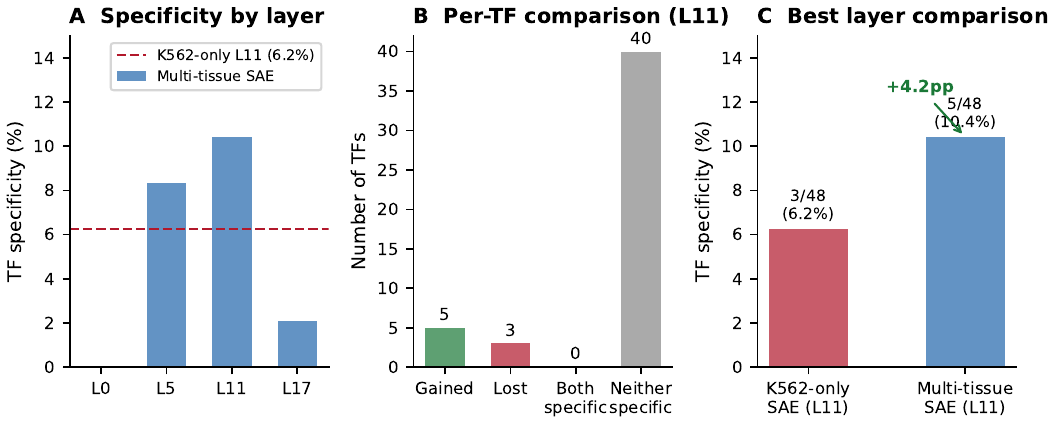}
\caption{\textbf{Multi-tissue SAE yields marginal improvement (6.2\%$\to$10.4\%).} Non-systematic gains establish Geneformer's representations as the bottleneck.}
\label{fig:multitissue}
\end{figure}

Table~\ref{tab:perturbation_comparison} presents the critical comparison (Fig.~\ref{fig:multitissue}). The best multi-tissue layer (L11) achieved \textbf{10.4\% TF specificity} (5/48 TFs)---an improvement of 4.2 percentage points over the K562-only result. However, the improvement fails three tests of systematicity:

\begin{enumerate}
\item \textbf{Non-systematic per-TF pattern.} Five TFs gained specificity (ATF5, BRCA1, GATA1, RBMX, NFRKB), three lost it (MAX, PHB2, SRF), and 40 were unchanged (Additional file~1: Table~S8). The sets are largely disjoint, suggesting stochastic variation.
\item \textbf{TF feature representation decreased.} The K562-only SAE has \emph{more} TF-associated features (64.5\%) than the multi-tissue SAE (60.5\%; Additional file~1: Table~S9).
\item \textbf{Layer pattern is informative.} L0 shows 0\% specificity, L5 shows 8.3\%, L11 peaks at 10.4\%, and L17 drops to 2.1\%---late-layer features are too abstract for regulatory mapping.
\end{enumerate}

\begin{table}[t]
\centering
\caption{\textbf{K562-only vs.\ multi-tissue SAE perturbation specificity.} Specificity = TRRUST TFs (48 total) with target-specific responding features.}
\label{tab:perturbation_comparison}
\smallskip
\begin{tabular}{llcc}
\toprule
SAE Training Data & Layer & TFs Specific & Rate \\
\midrule
K562-only & 11 & 3/48 & 6.2\% \\
\midrule
K562 + Tabula Sapiens & 0  & 0/48 & 0.0\% \\
K562 + Tabula Sapiens & 5  & 4/48 & 8.3\% \\
\textbf{K562 + Tabula Sapiens} & \textbf{11} & \textbf{5/48} & \textbf{10.4\%} \\
K562 + Tabula Sapiens & 17 & 1/48 & 2.1\% \\
\bottomrule
\end{tabular}
\end{table}

\subsection{Unannotated features: co-activation evidence and its limits}
\label{sec:unannotated}

A concern with the atlas is that 41--55\% of features lack ontology annotations. To investigate whether these are noise or encode biology not captured by existing databases, we applied two tests across four representative layers (0, 5, 11, 17).

First, we clustered unannotated features by Jaccard similarity of their top-20 gene sets. Only 2--3\% form standalone gene-set clusters, including ribosomal protein programs and mitochondrial complex assembly. Second, we tested guilt-by-association: whether unannotated features belong to co-activation modules alongside annotated features. The vast majority (95--98.5\%) of unannotated features co-activate with annotated features in biological modules (Additional file~1: Table~S10). Only 28--103 features per layer (1.4--4.7\% of unannotated features) are truly isolated.

However, this co-activation evidence has important limitations: (1)~with only 6--12 modules covering 96--99.5\% of all features, co-membership may be nearly unavoidable; (2)~features may co-activate due to shared statistical properties rather than shared biology; and (3)~no perturbation validation was performed. We interpret the high co-activation rate as evidence that most unannotated features are not random noise, while noting that the gap between ``not noise'' and ``biologically meaningful'' remains to be bridged.

\subsection{Success criteria scorecard}
\label{sec:scorecard}

Table~\ref{tab:scorecard} summarizes the success criteria defined for this study. Three of five criteria were exceeded, one was partially met (unannotated features participate in co-activation modules but form few standalone clusters), and one was not met. The unmet criterion---perturbation response specificity---is the scientifically most important: it establishes the boundary of what this model knows.

\begin{table}[t]
\centering
\caption{\textbf{Success criteria scorecard.} The unmet criterion (perturbation specificity) is the paper's central finding.}
\label{tab:scorecard}
\smallskip
\begin{tabular}{p{3.5cm}p{2.5cm}p{4cm}c}
\toprule
Criterion & Target & Result & Status \\
\midrule
Modules match pathways & $\geq$20 total & 141 (6--12/layer) & Exceeded \\
Causal specificity & Majority $>$2$\times$ & 60\% $>$2$\times$, median 2.36$\times$ & Exceeded \\
Perturbation specificity & $\geq$30\% & 6.2\% (3/48 TFs) & \textbf{Not met} \\
Unannotated structure & $\geq$10\% in clusters & 95--98.5\% in modules$^*$ & Partial \\
Cross-layer tracking & $\geq$50\% & 97--99.8\% highways & Exceeded \\
\bottomrule
\multicolumn{4}{l}{\small $^*$Co-activation module membership, not standalone clusters (2--3\%).}
\end{tabular}
\end{table}

\subsection{Feature space geometry}
\label{sec:geometry}

To visualize SAE feature organization, we projected decoder weight vectors using UMAP with cosine distance. All projections produced structureless point clouds. Quantitative analysis confirmed the cause: SAE decoder vectors are quasi-orthogonal by design (mean pairwise cosine similarity = 0.0007; within-module vs.\ between-module Cohen's $d = 0.075$). This quasi-orthogonality is a geometric signature of superposition: 4,608 features pack into 1,152 dimensions by spreading nearly uniformly across the hypersphere.

We therefore visualized the co-activation network directly using force-directed graph layout (Additional file~1: Figs.~S2--S4). Each module forms a spatially coherent community, with annotated features distributed across all communities while sparsely annotated features concentrate centrally. Annotation-based projections (TF-IDF weighted ontology vectors) provide independent validation that module structure reflects genuine biological similarity.

\subsection{Cell type enrichment mapping}
\label{sec:celltypes}

To connect SAE features to cellular identity, we performed cell type enrichment analysis across all layers of both models. For each feature, we computed its mean activation in cells of each cell type (among the 3,000 Tabula Sapiens cells spanning 56 cell types across immune, kidney, and lung tissues) and tested for enrichment using Fisher's exact test with Benjamini--Hochberg correction.

In the scGPT atlas, 2,028/2,048 features (99.0\%) at layer~7 are enriched for at least one cell type. Similar coverage was observed across all layers. The enrichment patterns are tissue-coherent: immune features activate preferentially in T cells, B cells, macrophages, and dendritic cells; kidney features in proximal tubule cells and podocytes; lung features in alveolar epithelial cells and pulmonary endothelial cells.

For the Geneformer atlas, we extracted Tabula Sapiens activations through Geneformer and encoded them with the K562-trained SAEs. Despite the SAEs being trained on K562 activations, they generalize to diverse tissue contexts. Both models' features tile cell identity space comprehensively, consistent with the models' training on diverse cell populations. These enrichments are available for interactive exploration in both web atlases.

\subsection{Interactive feature atlases}
\label{sec:atlases}

To enable community access to the complete feature atlases, we developed and deployed two interactive web platforms:

\begin{itemize}
\item \textbf{Geneformer Feature Atlas} (\url{https://biodyn-ai.github.io/geneformer-atlas/}): 82,525 features across 18 layers.
\item \textbf{scGPT Feature Atlas} (\url{https://biodyn-ai.github.io/scgpt-atlas/}): 24,527 features across 12 layers.
\end{itemize}

Both platforms provide six integrated views: Layer Explorer, Feature Detail Pages, Module Explorer, Cross-Layer Flow, Gene Search, and Ontology Search. The atlases are built with React, Vite, and Plotly.js, served via GitHub Pages, and require no installation.

\section{Discussion}

We have presented the first systematic application of sparse autoencoders to single-cell foundation models, producing comprehensive feature atlases for two architecturally distinct models: 82,525 features across 18 layers of Geneformer V2-316M and 24,527 features across 12 layers of scGPT whole-human. Both atlases reveal models that have organized biological knowledge into rich, modular, and interconnected internal representations---but representations that fundamentally encode \emph{co-expression and pathway structure} rather than \emph{causal regulatory logic}.

\paragraph{Superposition is massive and biologically productive.}
The finding that 99.8\% of SAE features are invisible to SVD, yet these novel features carry 98.7\% of all ontology annotations, has important implications for how we study neural network representations in biology. Standard dimensionality reduction techniques applied to model activations will miss the vast majority of encoded biological structure. The model uses its 1,152 dimensions to represent at least 82,525 distinct features through superposition---a compression ratio exceeding 70$\times$. Claims about what a model ``knows'' based on linear probing of its activation space may substantially underestimate the richness of learned representations.

\paragraph{Cross-model convergence despite architectural divergence.}
The parallel analysis reveals that superposition and biological feature organization are not artifacts of a single architecture. Despite fundamental differences---rank-value vs.\ continuous-value encoding, 18 vs.\ 12 layers, $d{=}1{,}152$ vs.\ $d{=}512$, next-token vs.\ masked gene prediction---both models develop modular feature organization (5--12 modules/layer), rich ontology annotation (29--59\%), and cross-layer information highways. The convergence on similar co-activation module counts (5--7 for scGPT, 6--12 for Geneformer) despite 2.25$\times$ dictionary size difference implies that the number of discoverable biological programs is determined more by the biology than by the model.

\paragraph{Hierarchical biological abstraction across layers.}
The U-shaped annotation profile, the shift in module themes from molecular machinery to integrative programs, and the complete representational turnover between layers together paint a picture of hierarchical biological abstraction. This parallels findings in large language models, where early layers handle token-level processing and later layers handle more abstract semantic computation, but here the ``tokens'' are genes and the ``semantics'' are biological programs.

\paragraph{Feature-level structure versus component-level nulls.}
The causal patching results (median 2.36$\times$ specificity, top 114.5$\times$) stand in stark contrast to the companion study's finding that attention head and MLP layer ablation produces null behavioral effects~\cite{kendiukhov2025systematic}. This implies that the model's biological computations are encoded at the feature level---in specific directions within the residual stream---rather than being localized to individual attention heads or MLP layers.

\paragraph{The co-expression--regulation dichotomy persists at every level of analysis.}
Our results extend the companion study's findings from attention weights to the residual stream. Attention-derived edge scores show zero incremental predictive value; component-level ablation produces null behavioral effects; and SAE features show rich biological annotation (45--59\%) but only 6.2\% regulatory specificity. These are independent methods probing fundamentally different aspects of model computation, yet all identify the same boundary. The model's SAE features respond to the overall shift in expression profile, activating features that match the \emph{new cell state} rather than features that specifically encode the TF's regulatory program.

\paragraph{The multi-tissue control establishes the model as the bottleneck.}
The marginal improvement (6.2\% $\to$ 10.4\%) with non-systematic gains and losses, combined with decreased TF feature representation (64.5\% $\to$ 60.5\%), indicates that the limitation is in Geneformer's representations, not in the SAE methodology or training data.

\paragraph{Implications for foundation model training.}
Current scFM training objectives may inherently bias representations toward co-expression. Learning causal regulatory relationships would require training signals that distinguish cause from correlation, such as perturbation prediction objectives. Our analysis suggests such objectives would need to be incorporated during pre-training.

\paragraph{SAEs as a general interpretability framework for biological models.}
Independent of the regulatory question, our work establishes SAEs as a productive framework for scFM interpretability. The successful application of an identical pipeline to two architecturally distinct models---with qualitatively consistent results---demonstrates that these tools are applicable to any transformer-based biological model.

\paragraph{Limitations.}
Our analysis has several limitations. First, the perturbation specificity test relies on TRRUST, which captures only a fraction of true relationships. Second, SAE features with $k{=}32$ at 4$\times$ expansion represent one point in architecture space. Third, causal patching tests only single-feature ablation; combinatorial effects remain unexplored. Fourth, the multi-tissue SAE used a simple pooling strategy. Fifth, the scGPT causal patching used proxy expression values, likely underestimating true specificity. Sixth, perturbation response mapping was performed only on Geneformer.

\section{Conclusions}

We demonstrate that sparse autoencoders successfully decompose the residual streams of Geneformer and scGPT into biologically interpretable feature atlases, confirming massive superposition (99.8\% of features invisible to SVD) and rich biological organization (29--59\% annotated, 141 co-activation modules, 97--99.8\% information highway coverage). However, these models encode co-expression and pathway structure rather than causal regulatory logic, as shown by the low perturbation specificity (6.2\%, confirmed as a model-level limitation by the multi-tissue control). These results define the boundary between what current single-cell foundation models know and what they do not, and point toward perturbation-aware training objectives as necessary for encoding regulatory relationships. The interactive feature atlases released with this work provide a new lens for understanding biological computation in transformer models.

\section{Methods}

\subsection{Models and data}

\paragraph{Geneformer.} We used Geneformer V2-316M~\cite{theodoris2023transfer} (18 transformer layers, 1,152 hidden dimensions, 18 attention heads per layer) from HuggingFace (\texttt{ctheodoris/\allowbreak Geneformer}, subfolder \texttt{Geneformer-V2-316M}). K562 data was obtained from the Replogle genome-scale CRISPRi dataset~\cite{replogle2022mapping}. We extracted 2,000 control cells (mean 2,028 genes per cell, 4,056,351 total token positions). For the multi-tissue experiment, we additionally used 3,000 cells from Tabula Sapiens~\cite{tabula2022tabula} (1,000 immune cells across 43 cell types, 1,000 kidney cells, 1,000 lung cells), yielding 5,623,164 token positions.

\paragraph{scGPT.} We used scGPT whole-human~\cite{cui2024scgpt} (12 transformer layers, 512 hidden dimensions, 8 attention heads per layer), trained on approximately 33 million single-cell transcriptomic profiles. Unlike Geneformer's rank-value tokenization, scGPT uses continuous expression values as input alongside gene token IDs from a vocabulary of approximately 60,700 genes. Genes are sorted by expression level (descending) and padded/truncated to a maximum sequence length of 1,200. Activations were extracted from the same 3,000 Tabula Sapiens cells (1,000 immune, 1,000 kidney, 1,000 lung), yielding 3,561,832 token positions per layer.

\subsection{Activation extraction}

For both models, we performed forward passes with PyTorch hooks registered at each transformer layer's output (after the residual connection), collecting hidden states at every gene position. Activations were stored as memory-mapped NumPy arrays (float32). Extraction used Apple Silicon MPS acceleration with batch size~1.

For Geneformer ($d{=}1{,}152$): 336.4~GB for 18 K562 layers (18.7~GB per layer); 103.6~GB for four Tabula Sapiens layers. For scGPT ($d{=}512$): approximately 82~GB for 12 layers ($\sim$6.8~GB per layer; variable due to per-cell sequence length differences).

\subsection{TopK sparse autoencoder architecture}

We used TopK SAEs~\cite{makhzani2013ksparse} with an architecture parameterized by input dimension $d$. The encoder maps $\mathbf{x} \in \mathbb{R}^{d}$ (centered by subtracting the training-set mean) through a linear projection $\mathbf{h} = W_\text{enc}(\mathbf{x} - \boldsymbol{\mu}) + \mathbf{b}_\text{enc}$ to a pre-activation vector $\mathbf{h} \in \mathbb{R}^{4d}$, followed by TopK sparsification retaining only the $k{=}32$ largest activations. The decoder reconstructs via $\hat{\mathbf{x}} = W_\text{dec}\mathbf{h}_\text{sparse} + \boldsymbol{\mu}$, where $W_\text{dec}$ columns are unit-normalized after each gradient step. Training minimized MSE: $\mathcal{L} = \|\mathbf{x} - \hat{\mathbf{x}}\|^2$.

For Geneformer ($d{=}1{,}152$): 4,608 features per layer, 1M subsampled positions per layer. For scGPT ($d{=}512$): 2,048 features per layer, trained on all 3,561,832 positions per layer. Both used identical hyperparameters: Adam optimizer, learning rate $3 \times 10^{-4}$, batch size 4,096, 5 epochs.

\subsection{Feature analysis and ontology annotation}

For each alive feature (activation frequency $> 0$ on 100K held-out positions), we identified the top 20 genes by mean activation magnitude. Each feature was tested for enrichment against: Gene Ontology Biological Process~\cite{ashburner2000go}, KEGG pathways~\cite{kanehisa2000kegg}, Reactome pathways~\cite{jassal2020reactome}, STRING protein--protein interactions~\cite{szklarczyk2023string}, and TRRUST TF--target relationships~\cite{han2018trrust}. Enrichment: Fisher's exact test (one-sided), Benjamini--Hochberg FDR at $\alpha = 0.05$. SVD comparison used top-50 singular vectors; ``SVD-aligned'' = decoder cosine $> 0.7$ with any SVD axis.

\subsection{Co-activation graph and module detection}

For each layer, PMI was computed between all pairs of alive features across 4,056,351 positions. Feature $i$ is ``active'' at position $j$ if among the top-$k$ ($k{=}32$) activations. $\text{PMI}(i,j) = \log_2 P(i,j) / [P(i)P(j)]$. Edges retained at PMI exceeding a significance threshold ($p < 0.001$ under permutation null). Community detection: Leiden algorithm~\cite{traag2019louvain}, resolution = 1.0.

\subsection{Causal feature patching}

For each of 50 annotated features at layer~11, single-feature ablation was performed using PyTorch forward hooks. At layer~11's output, the hidden state was encoded through the SAE, the target feature zeroed, decoded back, and the original hidden state replaced. The forward pass continued through layers 12--17 to produce output logits. Specificity ratio = $|\bar{\Delta}_\text{target}| / |\bar{\Delta}_\text{other}|$. Each feature tested on 200 cells.

\subsection{Perturbation response mapping}

For each of 100 CRISPRi targets (48 TRRUST TFs, 52 other genes), Geneformer activations were extracted from 20 perturbed cells and encoded through the layer-11 SAE. Feature responses identified by Wilcoxon rank-sum test against 100K control positions (BH correction, $|\text{effect}| > 0.5$). For TRRUST TFs, enrichment of responding features' top genes for known targets: Fisher's exact test.

\subsection{Cross-layer computational graph}

PMI between SAE feature activations at source and target layers, encoding the same positions through both layer's SAEs. For Geneformer: three layer pairs (L0$\to$L5, L5$\to$L11, L11$\to$L17), 500,000 positions each. For scGPT: three layer pairs (L0$\to$L4, L4$\to$L8, L8$\to$L11), all 3,561,832 positions. Information highway = source feature with $\geq$1 target feature at PMI~$> 3$.

\subsection{Multi-tissue SAE}

Multi-tissue SAEs trained on 500K K562 + 500K Tabula Sapiens positions per layer (balanced 1M pool). Architecture, training, and hyperparameters identical to K562-only SAEs. Trained at layers 0, 5, 11, 17. Feature analysis, annotation, and perturbation mapping used same protocols.

\subsection{Novel feature characterization}

Unannotated features at layers 0, 5, 11, 17 were clustered by Jaccard similarity of top-20 gene sets (Leiden, resolution = 0.5). Guilt-by-association: co-membership in co-activation modules with annotated features.

\subsection{Cell type enrichment analysis}

For each feature at each layer, we computed the mean activation magnitude in cells belonging to each of 56 cell types from the Tabula Sapiens dataset. Enrichment assessed by Fisher's exact test (BH-corrected FDR~$< 0.05$).

\subsection{Interactive web atlases}

Both feature atlases were deployed as single-page web applications built with React 18, TypeScript, Vite 6, Tailwind CSS, and Plotly.js. All feature data were preprocessed into compact JSON files served statically via GitHub Pages. Source code: \url{https://github.com/Biodyn-AI/geneformer-atlas} and \url{https://github.com/Biodyn-AI/scgpt-atlas}.

\subsection{Dimensionality reduction and visualization}

We projected decoder weight vectors to 2D using UMAP~\cite{mcinnes2018umap} with cosine distance, obtaining structureless point clouds due to quasi-orthogonality (mean pairwise cosine = 0.0007). We therefore adopted force-directed graph layout (Fruchterman--Reingold via NetworkX) of intra-module co-activation edges. For independent validation, annotation-based feature vectors (TF-IDF weighted ontology terms) were projected using UMAP ($n_\text{neighbors}=15$, $\min_\text{dist}=0.1$, cosine metric) and t-SNE (perplexity~20, cosine metric).

\section*{Declarations}

\subsection*{Ethics approval and consent to participate}
Not applicable. This study used only publicly available pre-trained models and previously published datasets. No human participants were involved.

\subsection*{Consent for publication}
Not applicable.

\subsection*{Competing interests}
The author declares no competing interests.

\subsection*{Authors' contributions}
IK is the sole author of this work. IK conceptualized the study; developed the methodology including the sparse autoencoder training pipeline, causal patching framework, and cross-layer computational graph analysis; wrote all software for data extraction, model training, feature annotation, perturbation mapping, and visualization; performed all computational experiments across both Geneformer and scGPT models; built the two interactive web atlases; performed the formal analysis; interpreted the results; and wrote, reviewed, and approved the final manuscript.

\subsection*{Funding}
No external funding was received for this work.

\subsection*{Availability of data and materials}
All analysis code, trained SAE models, and feature catalogs are available at \url{https://github.com/Biodyn-AI/bio-sae}. Interactive feature atlases: Geneformer (\url{https://biodyn-ai.github.io/geneformer-atlas/}; source: \url{https://github.com/Biodyn-AI/geneformer-atlas}) and scGPT (\url{https://biodyn-ai.github.io/scgpt-atlas/}; source: \url{https://github.com/Biodyn-AI/scgpt-atlas}). Geneformer V2-316M: HuggingFace \texttt{ctheodoris/Geneformer}. scGPT whole-human: available from the scGPT authors~\cite{cui2024scgpt}. Replogle CRISPRi data~\cite{replogle2022mapping}. Tabula Sapiens~\cite{tabula2022tabula}. TRRUST~\cite{han2018trrust}.

\subsection*{Acknowledgements}
Computations were performed on Apple Silicon hardware with MPS acceleration.

\bibliography{references}

\clearpage
\appendix
\renewcommand{\thetable}{S\arabic{table}}
\renewcommand{\thefigure}{S\arabic{figure}}
\setcounter{table}{0}
\setcounter{figure}{0}

\section*{Supplementary Material}


\begin{table}[H]
\centering
\caption{\textbf{Per-ontology enrichment counts across all 18 Geneformer layers.} Each entry is the number of significant enrichments (FDR $< 0.05$) for features at that layer. GO~BP = Gene Ontology Biological Process. TRRUST columns show enrichment for TF target sets (TF) and individual TF$\to$target edges (Edges).}
\label{tab:perontology}
\smallskip
\begin{tabular}{rrrrrrc}
\toprule
Layer & GO BP & KEGG & Reactome & STRING & TRRUST TF & TRRUST Edges \\
\midrule
0  & 10,153 & 2,650 & 11,001 & 302 & 155 & 42 \\
1  & 10,022 & 2,433 & 10,512 & 258 & 164 & 48 \\
2  & 9,948  & 2,495 & 10,790 & 283 & 150 & 32 \\
3  & 9,726  & 2,514 & 9,525  & 273 & 157 & 37 \\
4  & 8,537  & 2,045 & 9,195  & 248 & 133 & 30 \\
5  & 7,695  & 1,845 & 8,189  & 216 & 136 & 24 \\
6  & 7,180  & 1,555 & 7,871  & 182 & 110 & 28 \\
7  & 6,628  & 1,637 & 7,080  & 181 & 125 & 30 \\
8  & 6,850  & 1,570 & 7,169  & 199 & 90  & 27 \\
9  & 7,299  & 1,643 & 7,880  & 207 & 103 & 29 \\
10 & 8,461  & 1,751 & 9,247  & 214 & 128 & 30 \\
11 & 8,785  & 2,089 & 8,957  & 227 & 112 & 31 \\
12 & 8,217  & 1,915 & 8,856  & 210 & 117 & 35 \\
13 & 7,158  & 1,686 & 8,393  & 202 & 101 & 28 \\
14 & 7,615  & 1,595 & 8,412  & 221 & 97  & 27 \\
15 & 6,790  & 1,520 & 7,135  & 150 & 126 & 34 \\
16 & 7,040  & 1,781 & 7,172  & 158 & 87  & 25 \\
17 & 7,002  & 1,762 & 6,869  & 193 & 131 & 25 \\
\midrule
\textbf{Total} & \textbf{145,106} & \textbf{34,486} & \textbf{154,253} & \textbf{3,924} & \textbf{2,222} & \textbf{562} \\
\bottomrule
\end{tabular}
\end{table}

\begin{table}[H]
\centering
\caption{\textbf{Cross-layer feature persistence from layer~0.} Matches = features at L0 with cosine similarity $> 0.7$ to any feature at the target layer. The model undergoes radical representational transformation: by layer~6, essentially all features are novel with no L0 ancestry.}
\label{tab:persistence}
\smallskip
\begin{tabular}{lcc}
\toprule
L0 $\to$ Target & Matches (cos $> 0.7$) & Rate \\
\midrule
L0 $\to$ L1  & 114 & 2.5\% \\
L0 $\to$ L2  & 93 & 2.0\% \\
L0 $\to$ L4  & 67 & 1.5\% \\
L0 $\to$ L6  & 25 & 0.5\% \\
L0 $\to$ L8  & 10 & 0.2\% \\
L0 $\to$ L10 & 1 & $\sim$0\% \\
L0 $\to$ L12+ & 0 & 0\% \\
\bottomrule
\end{tabular}
\end{table}

\begin{table}[H]
\centering
\caption{\textbf{Co-activation module statistics across all 18 layers.} PMI-based graphs with Leiden clustering (resolution = 1.0). Modules = number of distinct communities. Coverage = fraction of alive features in at least one module.}
\label{tab:modules_full}
\smallskip
\begin{tabular}{rccrc}
\toprule
Layer & Modules & Feats in Modules & PMI Edges & Coverage \\
\midrule
0  & 6  & 4,577 & 446,324 & 99.3\% \\
1  & 8  & 4,562 & 440,681 & 99.0\% \\
2  & 7  & 4,536 & 404,403 & 98.6\% \\
3  & 8  & 4,518 & 393,574 & 98.3\% \\
4  & 9  & 4,502 & 393,194 & 98.2\% \\
5  & 12 & 4,472 & 390,845 & 97.7\% \\
6  & 7  & 4,458 & 383,033 & 97.3\% \\
7  & 8  & 4,439 & 371,832 & 96.7\% \\
8  & 7  & 4,478 & 369,280 & 97.6\% \\
9  & 7  & 4,535 & 380,304 & 98.7\% \\
10 & 9  & 4,571 & 388,498 & 99.3\% \\
11 & 8  & 4,565 & 388,103 & 99.3\% \\
12 & 7  & 4,567 & 388,977 & 99.5\% \\
13 & 7  & 4,561 & 383,779 & 99.5\% \\
14 & 8  & 4,543 & 379,595 & 99.5\% \\
15 & 8  & 4,461 & 340,269 & 98.2\% \\
16 & 8  & 4,358 & 327,895 & 96.0\% \\
17 & 7  & 4,474 & 343,059 & 97.7\% \\
\midrule
\textbf{Total} & \textbf{141} & & & \\
\bottomrule
\end{tabular}
\end{table}

\begin{table}[H]
\centering
\caption{\textbf{Top 10 causally specific SAE features at layer~11.} $\Delta$Target and $\Delta$Other = mean logit change at target and off-target gene positions, respectively, upon zeroing the feature. Specificity ratios were computed from unrounded values; displayed $\Delta$ values are rounded to three decimal places.}
\label{tab:causal_top10}
\smallskip
\begin{tabular}{clccc}
\toprule
Feature & Annotation & Specificity & $\Delta$Target & $\Delta$Other \\
\midrule
F2035 & Cell Differentiation (neg.\ reg.) & 114.5$\times$ & $-$0.208 & $+$0.002 \\
F3692 & ERAD Pathway & 108.1$\times$ & $-$0.129 & $-$0.001 \\
F3933 & Intracellular Signaling (neg.\ reg.) & 55.7$\times$ & $-$0.196 & $-$0.004 \\
F157  & Golgi Vesicle Transport & 25.4$\times$ & $-$0.056 & $-$0.002 \\
F3532 & Protein Metabolic Process (pos.\ reg.) & 11.2$\times$ & $-$0.127 & $-$0.011 \\
F4516 & Mitotic Spindle Microtubules & 10.6$\times$ & $+$0.672 & $+$0.063 \\
F1337 & Cell Cycle Phase Transition & 9.4$\times$ & $-$0.058 & $-$0.006 \\
F1023 & Mitotic Spindle Microtubules & 7.6$\times$ & $-$2.799 & $-$0.367 \\
F2936 & Mitochondrion Organization & 7.1$\times$ & $-$0.366 & $-$0.051 \\
F3962 & Endocytosis & 6.9$\times$ & $-$0.099 & $-$0.014 \\
\bottomrule
\end{tabular}
\end{table}

\begin{table}[H]
\centering
\caption{\textbf{Geneformer cross-layer information highways.} PMI between SAE feature activations at source and target layers (500K positions each). A highway = source feature with $\geq$1 target-layer feature at PMI $> 3$.}
\label{tab:highways}
\smallskip
\begin{tabular}{lccccc}
\toprule
Layer Pair & Feats w/ Deps & Mean Max PMI & Median Max PMI & Max PMI & Highways \\
\midrule
L0 $\to$ L5   & 4,604 & 6.61 & 6.72 & 11.10 & 4,530 (98.4\%) \\
L5 $\to$ L11  & 4,518 & 6.63 & 6.71 & 10.87 & 4,401 (97.4\%) \\
L11 $\to$ L17 & 4,555 & 6.79 & 6.86 & 10.66 & 4,544 (99.8\%) \\
\bottomrule
\end{tabular}
\end{table}

\begin{table}[H]
\centering
\caption{\textbf{scGPT cross-layer information highways.} Same methodology as Table~\ref{tab:highways}. Note the progressive drop in downstream connectivity.}
\label{tab:scgpt_highways}
\smallskip
\begin{tabular}{lrccc}
\toprule
Layer Pair & PMI Edges & Upstream & Downstream & Max PMI \\
\midrule
L0 $\to$ L4  & 75,305 & 1,935/2,027 (95.5\%) & 1,960/2,048 (95.7\%) & 9.15 \\
L4 $\to$ L8  & 61,263 & 1,955/2,048 (95.5\%) & 1,723/2,048 (84.1\%) & 9.26 \\
L8 $\to$ L11 & 45,258 & 1,773/2,048 (86.6\%) & 1,289/2,048 (62.9\%) & 10.78 \\
\bottomrule
\end{tabular}
\end{table}

\begin{table}[H]
\centering
\caption{\textbf{Top cross-layer biological cascades.} Strongest annotated PMI connections between layer pairs. ``Unlabeled'' = the target feature lacks direct ontology annotation.}
\label{tab:cascades}
\smallskip
\small
\begin{tabular}{lp{2.6cm}p{2.6cm}p{3.2cm}r}
\toprule
Pair & Source Feature & Target Feature & Biological Logic & PMI \\
\midrule
L0$\to$L5  & Protein Processing in ER & \textit{unlabeled} & ER stress cascade & 11.10 \\
L0$\to$L5  & mTORC1 Regulation & Autophagy & mTORC1$\to$autophagy & 9.55 \\
L0$\to$L5  & Wnt Signaling & \textit{unlabeled} & Wnt pathway processing & 9.48 \\
\midrule
L5$\to$L11 & Protein Polyubiq. & \textit{unlabeled} & Protein quality control & 10.87 \\
L5$\to$L11 & Translation & \textit{unlabeled} & Translational regulation & 10.35 \\
L5$\to$L11 & RNA Splicing (neg.\ reg.) & \textit{unlabeled} & Post-transcriptional & 10.21 \\
\midrule
L11$\to$L17 & Protein Modification & Angiogenesis (pos.\ reg.) & PTM$\to$phenotype & 10.62 \\
L11$\to$L17 & COPII Vesicle Budding & Thermogenesis & Secretory$\to$metabolic & 10.29 \\
L11$\to$L17 & Actomyosin Org. & Cell Locomotion (neg.\ reg.) & Structure$\to$motility & 10.14 \\
\bottomrule
\end{tabular}
\end{table}

\begin{table}[H]
\centering
\caption{\textbf{Per-TF head-to-head comparison at layer~11.} Only TFs with changed specificity are shown; 40 additional TFs had no specific features in either condition.}
\label{tab:per_tf}
\smallskip
\begin{tabular}{lccc}
\toprule
TF & K562-SAE Specific & MT-SAE Specific & Change \\
\midrule
ATF5  & 0 & 1 & gained \\
BRCA1 & 0 & 1 & gained \\
GATA1 & 0 & 1 & gained \\
RBMX  & 0 & 1 & gained \\
NFRKB & 0 & 1 & gained \\
\midrule
MAX   & 1 & 0 & lost \\
PHB2  & 1 & 0 & lost \\
SRF   & 1 & 0 & lost \\
\bottomrule
\end{tabular}
\end{table}

\begin{table}[H]
\centering
\caption{\textbf{TF feature diagnostics.} Features with known TFs in top-20 genes and TF-dominant features ($\geq$3 TFs in top genes). Denominators = features with non-empty top-20 gene lists (may differ from alive counts because some features that are ``dead'' on the 100K held-out sample still produce gene lists from the full training data). K562-only SAE has more TF-associated features than the multi-tissue SAE.}
\label{tab:tf_diagnostics}
\smallskip
\begin{tabular}{lcc}
\toprule
SAE & Features with TFs in top genes & TF-dominant ($\geq$3 TFs) \\
\midrule
K562-only L11     & 2,967/4,598 (64.5\%) & 424 \\
Multi-tissue L0   & 2,796/4,608 (60.7\%) & 452 \\
Multi-tissue L5   & 2,777/4,568 (60.8\%) & 337 \\
Multi-tissue L11  & 2,782/4,601 (60.5\%) & 343 \\
Multi-tissue L17  & 2,680/4,603 (58.2\%) & 346 \\
\bottomrule
\end{tabular}
\end{table}

\begin{table}[H]
\centering
\caption{\textbf{Unannotated feature analysis.} Co-activate = unannotated feature belongs to a co-activation module containing annotated features. Isolated = no module membership. A small number of features (3 at L11, 17 at L17) belong to modules containing only unannotated features and are excluded from both columns. Clusters = standalone gene-set clusters among unannotated features.}
\label{tab:novel}
\smallskip
\begin{tabular}{rrrcrr}
\toprule
Layer & Annotated & Unannotated & Clusters & Co-activate w/ Annotated & Isolated \\
\midrule
0  & 2,702 & 1,906 & 15 (48 feats) & 1,876 (98.4\%) & 30 (1.6\%) \\
5  & 2,383 & 2,193 & 19 (69 feats) & 2,090 (95.3\%) & 103 (4.7\%) \\
11 & 2,583 & 2,015 & 11 (47 feats) & 1,984 (98.5\%) & 28 (1.4\%) \\
17 & 2,154 & 2,426 & 12 (58 feats) & 2,334 (96.2\%) & 75 (3.1\%) \\
\bottomrule
\end{tabular}
\end{table}

\clearpage


\begin{figure}[H]
\centering
\includegraphics[width=\textwidth]{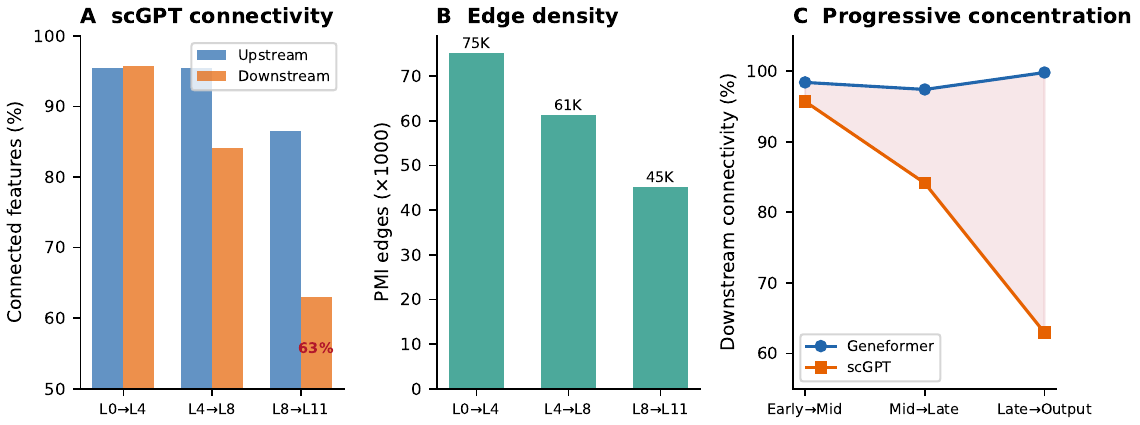}
\caption{\textbf{scGPT cross-layer connectivity reveals progressive information concentration.} \textbf{(A)}~Upstream connectivity remains high (86--96\%) but downstream connectivity drops sharply from 96\% to 63\% across layer pairs, indicating progressive bottlenecking. \textbf{(B)}~PMI edge density decreases from 75K to 45K edges across the three layer pairs. \textbf{(C)}~Comparison with Geneformer: Geneformer maintains near-complete downstream connectivity (97--100\%) while scGPT drops to 63\%, suggesting fundamentally different information flow architectures.}
\label{fig:scgpt_highways}
\end{figure}

\begin{figure}[H]
\centering
\includegraphics[width=\textwidth,height=0.85\textheight,keepaspectratio]{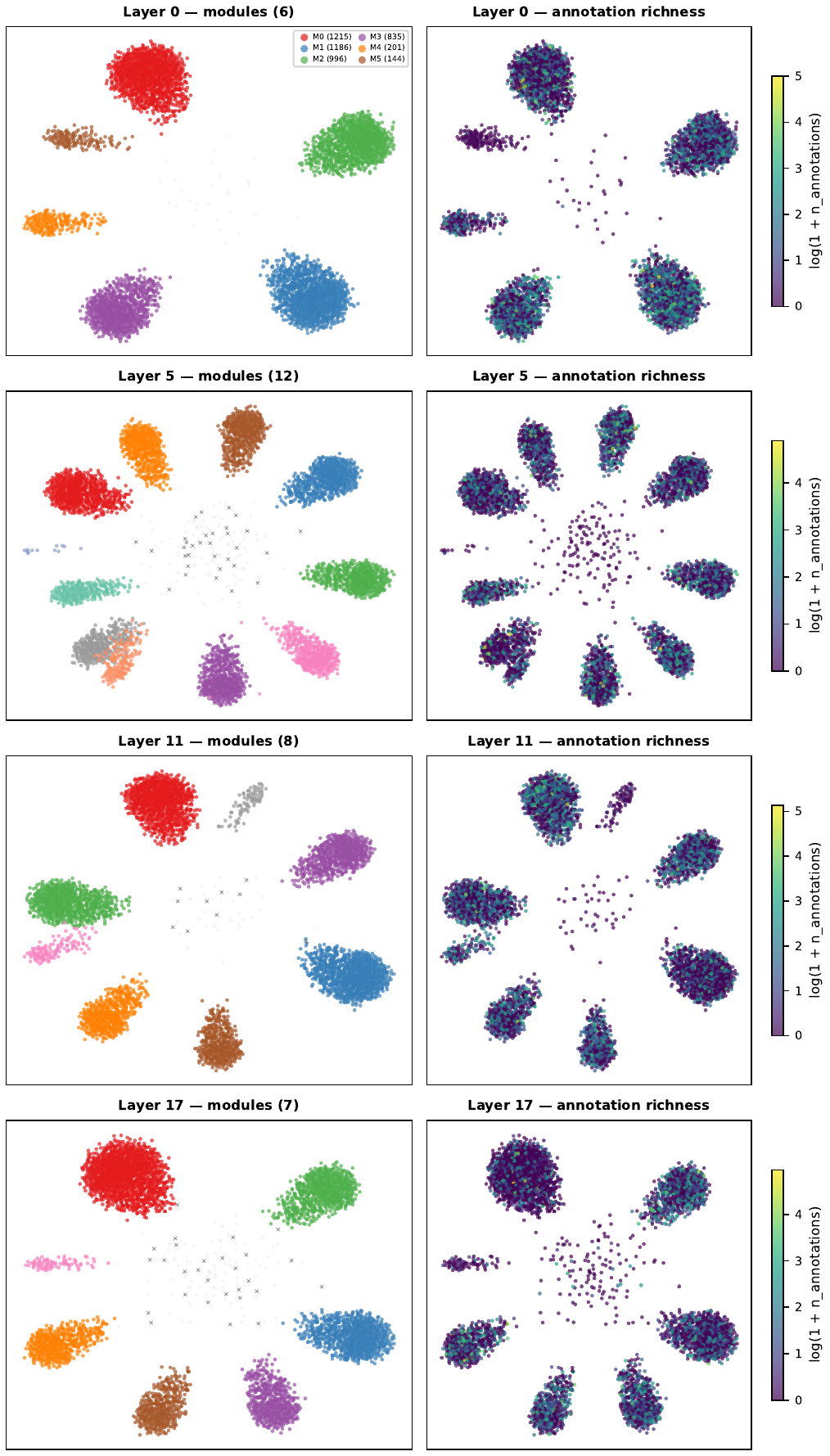}
\caption{\textbf{Co-activation network layout of SAE features across layers.} Left column: force-directed graph layout of intra-module co-activation edges, colored by Leiden module membership. Each module forms a spatially distinct community. Right column: same layouts colored by annotation richness (log-transformed number of significant enrichment terms). Unassigned features (gray) cluster centrally. Module count varies from 6 (L0) to 12 (L5), reflecting the complexity of co-activation patterns at different layers.}
\label{fig:umap_overview}
\end{figure}

\begin{figure}[H]
\centering
\includegraphics[width=\textwidth]{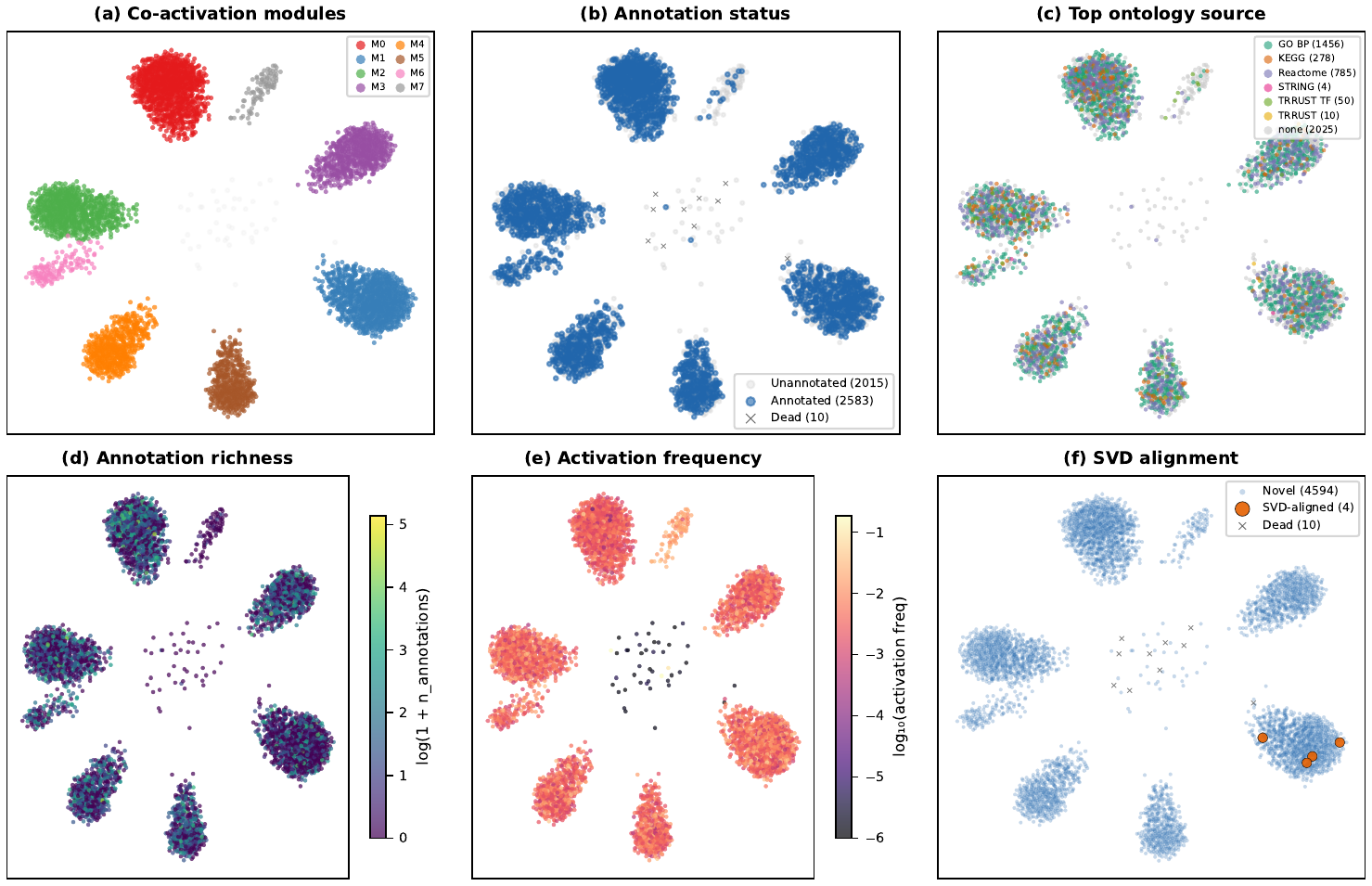}
\caption{\textbf{Six-panel co-activation layout of layer~11 SAE features.} \textbf{(a)}~Eight Leiden modules form spatially distinct communities. \textbf{(b)}~Annotated features (blue) distribute across all module clusters; unannotated features (gray) concentrate centrally. \textbf{(c)}~Top ontology source reveals module-specific enrichment patterns: certain modules are dominated by GO~BP (green), others by STRING interactions (pink) or Reactome pathways (purple). \textbf{(d)}~Annotation richness gradient across modules. \textbf{(e)}~Activation frequency varies systematically across modules. \textbf{(f)}~SVD-aligned features (orange, $n=4$) are scattered across different modules, while 4,594 novel features (blue) fill the landscape.}
\label{fig:l11_detail}
\end{figure}

\begin{figure}[H]
\centering
\includegraphics[width=\textwidth]{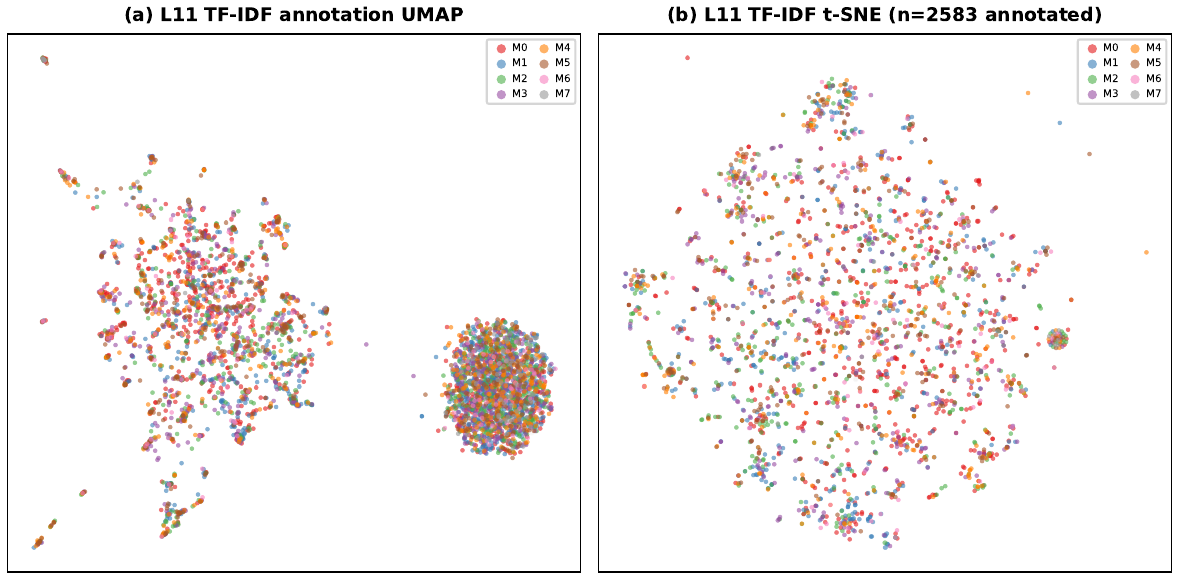}
\caption{\textbf{Annotation-based projections provide independent validation.} \textbf{(a)}~UMAP of TF-IDF weighted ontology term vectors for layer~11 (4,608 features). Annotated features (left cluster) separate from unannotated features (right blob), with internal structure reflecting shared biological annotations. \textbf{(b)}~t-SNE of TF-IDF annotation vectors for annotated features only ($n=2{,}583$). Fine-grained subclusters partially correspond to co-activation modules, confirming that module structure reflects genuine biological similarity.}
\label{fig:cross_layer_umap}
\end{figure}

\end{document}